\documentclass[9pt]{article}
\usepackage[utf8]{inputenc}
\usepackage[margin=1in]{geometry}
\usepackage{authblk}
\usepackage{hyperref}
\usepackage{amsmath}
\usepackage{mathtools}
\usepackage{amssymb}
\usepackage{graphicx}
\usepackage{xcolor}
\usepackage{comment}
\usepackage{booktabs}
\usepackage{multirow}
\usepackage{pdflscape}
\usepackage[super]{nth}
\usepackage{fancyhdr}

\newif\ifhighlights

\ifhighlights
    \newcommand{\textaddition}[1]{\textcolor{blue}{#1}}
    \newcommand{\textcorrection}[1]{\textcolor{red}{#1}}

\else
    \newcommand{\textaddition}[1]{#1}
    \newcommand{\textcorrection}[1]{#1}

\fi

\fancypagestyle{titlepagestyle}{
  \fancyhead{} 
  \fancyhead[R]{FERMILAB-FN-1291-AD} 
}

\title{Proton Dynamics Scenarios in the
Integrable Optics Test Accelerator (IOTA) at Fermilab}
\author{N. Banerjee\thanks{nilanjan@fnal.gov}, A. Romanov, G. Stancari, M. Wallbank}
\affil{Fermi National Accelerator Laboratory, Batavia, IL}
\date{November 22, 2025}

\begin{document}

\maketitle

\thispagestyle{titlepagestyle}

\abstract{The Integrable Optics Test Accelerator (IOTA) at Fermilab provides a versatile platform for studying the interplay of space-charge, impedance, and non-linear optics in high-intensity hadron beams within synchrotrons and storage rings. This report examines the parameters and dynamics of 2.5~MeV proton beam operations in two configurations of the bare IOTA lattice \textaddition{(dipoles, quadrupoles, sextupoles, and rf cavity only)}: one for demonstrating Non-linear Integrable Optics with the Danilov-Nagaitsev magnet, and the other for use with electron cooling. We offer order-of-magnitude estimates of the transverse emittance growth rate as a function of beam intensity, highlighting contributions from residual gas scattering, intra-beam scattering, and space-charge effects. Under nominal conditions, the beam lifetime is projected to be less than 7~minutes at low intensity with the current vacuum quality, and fewer than 100,000~turns at high intensity due to strong space-charge effects. The calculations presented here will guide strategies to mitigate emittance growth and inform future IOTA experiments.}

\section{Introduction}

\begin{figure}
    \centering
    \includegraphics[width=0.8\columnwidth]{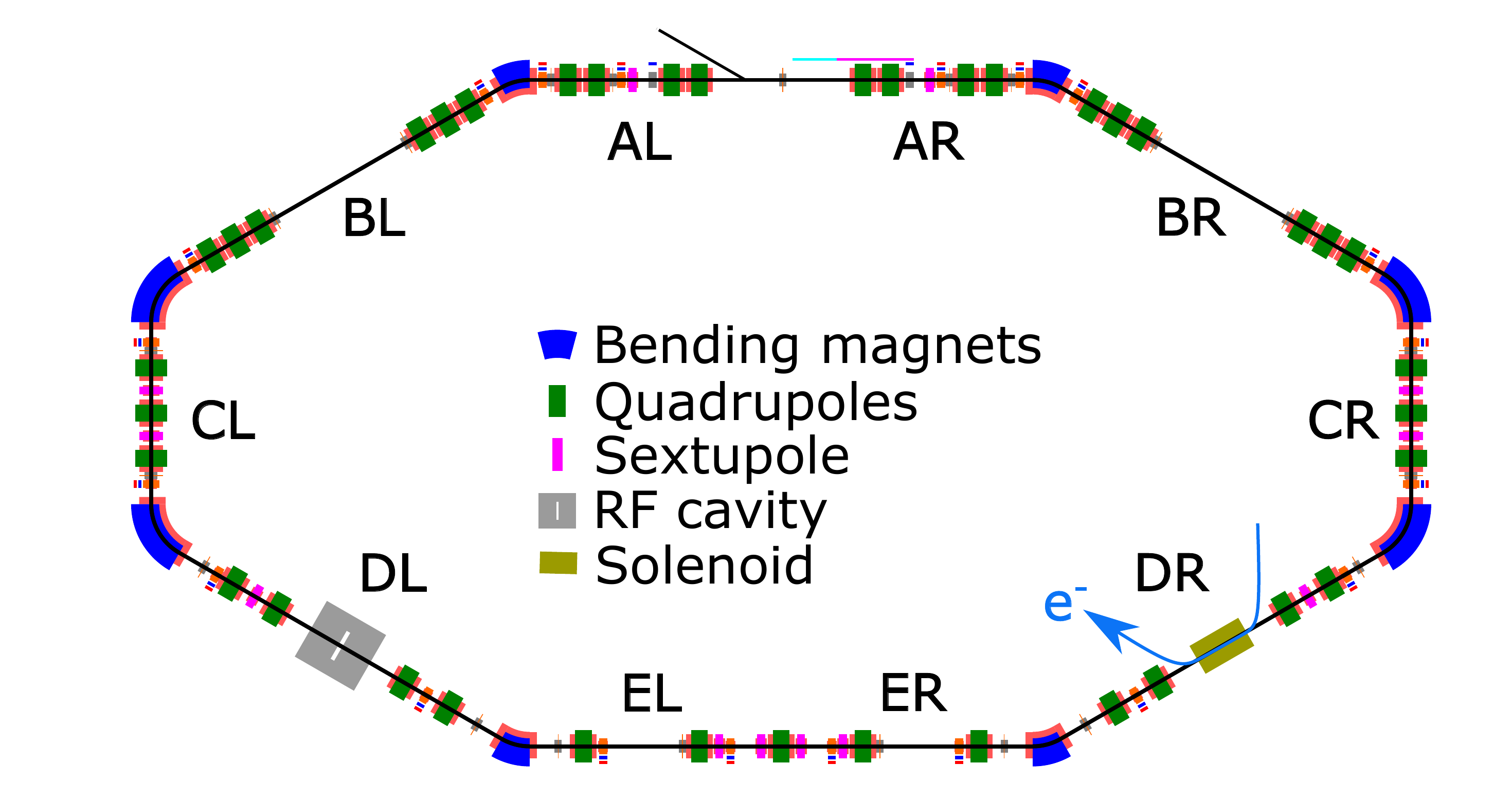}
    \caption{Layout of the Integrable Optics Test Accelerator \cite{Antipov2017} without special non-linear magnets in sections BL and BR. The beam moves clockwise. The blue arrow indicates the propagation of electrons in the proposed cooler to be situated in section DR.}
    \label{fig:iota}
\end{figure}

Future proton drivers for neutron \cite{Galambos2020}, neutrino \cite{Garoby2016}, and muon generation \cite{Boscolo2019} aim to deliver beams with intensity and phase-space density an order of magnitude greater than current achievements. At Fermilab, we have proposed the Accelerator Complex Evolution plan \cite{Ainsworth2023} to significantly increase the number of protons delivered to the Long-Baseline Neutrino Facility (LBNF) target compared to \textcorrection{what is expected from the PIP-II upgrade} \cite{Lebedev2017}. A key component of this plan is replacing the Fermilab Booster \cite{Eldred2021}, possibly with a new Rapid Cycling Synchrotron (RCS) capable of operating at high intensity and managing high space-charge tune shifts. To tackle emittance growth, beam loss, and instabilities in this high-intensity regime, it is crucial to advance our understanding of space-charge compensation, the interaction between coherent instabilities and space-charge \cite{Burov2018, Burov2019, Buffat2021}, and the best practices for applying Landau damping while maintaining dynamic aperture. The proton program at the Integrable Optics Test Accelerator (IOTA) \cite{Antipov2017} employs 2.5~MeV (with $pc \approx 68.5$~MeV/c) protons to mimic beams with incoherent betatron tune-shifts approaching 0.5 in a controlled impedance environment, utilizing a digital wake-building feedback system known as the \textit{waker} \cite{Ainsworth2021} (see Fig.~\ref{fig:iota} for the layout of the \textcorrection{bare storage ring including dipoles, quadrupoles, sextupoles, and the rf cavity}). IOTA enables a wide range of experiments on Non-linear Integrable Optics \cite{Valishev2021, Wieland2023}, space-charge compensation \cite{Shiltsev2017}, electron lens \cite{Stancari2021}, instabilities, and Landau damping \cite{Antipov2021}. The knowledge gained will contribute to the design of future hadron synchrotrons and storage rings.\\

While the low-energy proton beam \textcorrection{will provide} a valuable platform for isolating the dynamics of intense space-charge from coherent instabilities, it \textcorrection{will be} vulnerable to significant emittance growth and loss mechanisms. \textcorrection{Estimates of single Coulomb scattering of protons by residual gas nuclei suggest that beam lifetime will be limited through large-angle scattering events, whereas multiple inelastic scattering with atomic electrons will cause emittance growth.} A 2.5~MeV proton beam with an rms emittance of 4~\textmu m, circulating in IOTA with a nominal beta function of 4~m (maximum 10~m) and a chamber diameter of 50~mm, \textcorrection{will experience} emittance growth and beam loss on time scales of 70~seconds and 16~minutes, respectively, due to residual gas scattering at an effective residual gas density of $7.0\times 10^8$~cm\textsuperscript{-3}, as measured in atomic hydrogen equivalent using the dynamics of single electrons \cite{Romanov2021} in IOTA. Additionally, intra-beam scattering (IBS) \textcorrection{will result} in faster emittance growth, with a time scale of a few seconds at the phase-space densities required for studies involving intense space-charge. Therefore, sustaining the beam in IOTA will require compensating for heating mechanisms through cooling. In this report, we explore the expected range of emittance growth rates of the proton beam through various mechanisms, which will guide the commissioning of protons in the IOTA ring, procedures for planned experiments, and the design of the electron cooler \cite{Budker1967} as part of the broader electron-lens research program.\\

In the next section, we describe the full range of parameters for the beam generated by the IOTA proton injector. Following that, we discuss two linear optics configurations and derive the ranges of ring acceptance, longitudinal dynamics parameters, and beam intensity. Next, we calculate the range of emittance growth and beam loss caused by elastic scattering processes and space-charge effects. Finally, we discuss the implications of our calculations for conducting experiments with 2.5~MeV protons.\\

\section{Beam from the IOTA Proton Injector}

\begin{table}
    \caption{Beam parameters for injection into IOTA.}
    \label{tab:protoninj}
    \centering
    \begin{tabular}{lccl}
        \toprule
        \textbf{Parameter} & \textbf{Nominal} & \textbf{Range} & \textbf{Unit} \\
        \midrule
        Kinetic energy ($K_p$) & 2.5 & $2.3 - 2.5$ & MeV \\
        \textaddition{Geometric} emittances ($\epsilon_x, \epsilon_y$) & 4.3, 3 & $3.0 - 9.0$ & \textmu m \\
        RMS momentum spread ($\sigma_{\delta,\mathrm{inj}}$) & 1.32 & $1-2$ & $10^{-3}$\\
        Bunch Length ($\sigma_{z,\mathrm{inj}}$) & 6.6 & & mm \\
        Beam current ($I_p$) & --- & $\leq 8$ & mA \\
        \bottomrule
    \end{tabular}
\end{table}

\begin{figure}
    \centering
    \includegraphics[width=0.5\columnwidth]{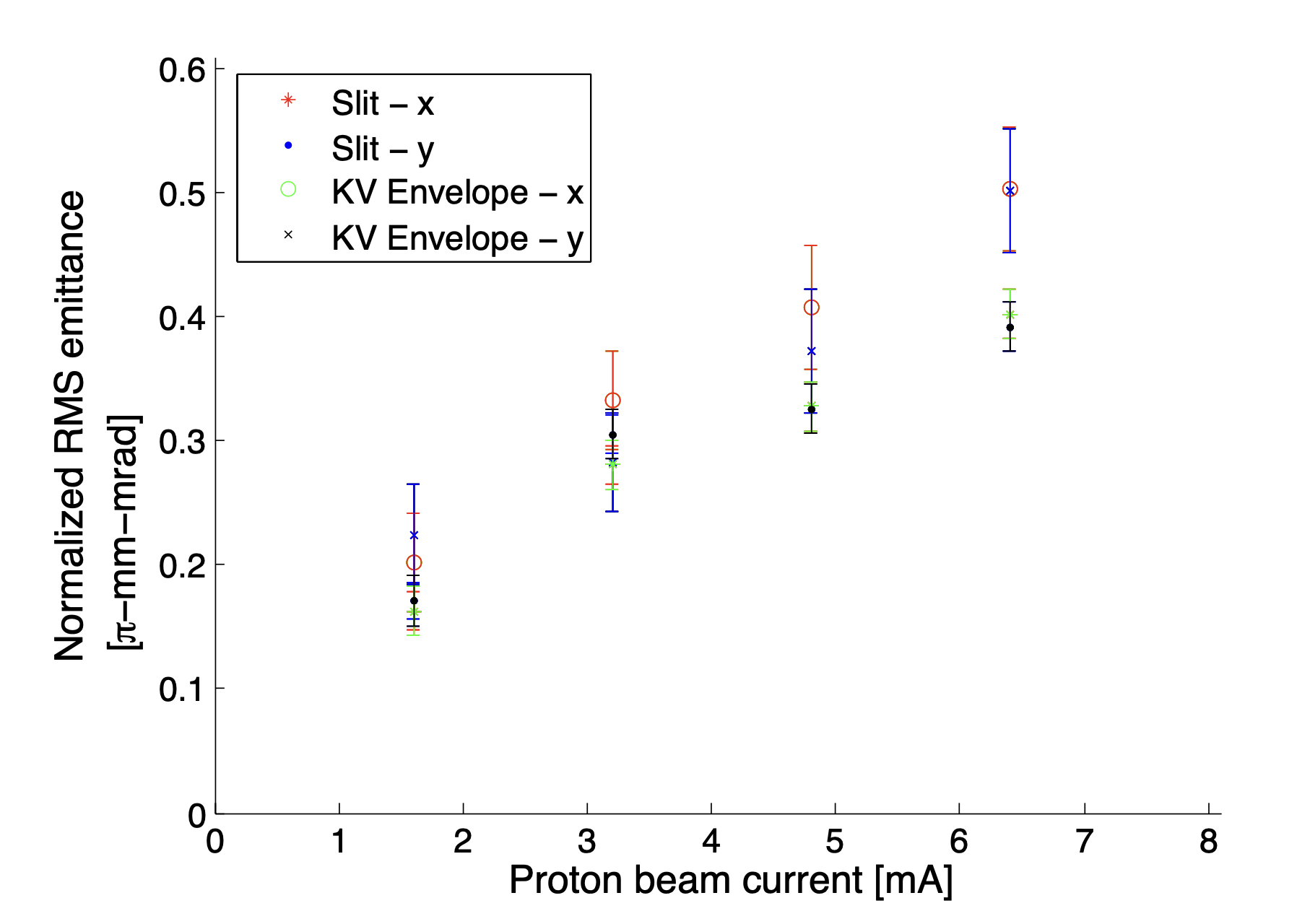}
    \caption{Normalized rms emittance of protons as a function of beam current as measured in a previous test of the duoplasmatron source and the Low Energy Beam Transport section \textcorrection{Reproduced with permission from \cite{Tam2010}.}}
    \label{fig:ipilebt}
\end{figure}

The proton beam parameters in IOTA depend on the configurations of both the injector and the ring lattice. The IOTA Proton Injector (IPI) \cite{Edstrom2023} consists of a duoplasmatron source that generates protons at 50~keV, followed by a Low Energy Beam Transport (LEBT) section to deliver the beam to the Radio Frequency Quadrupole (RFQ) \cite{Ostroumov2006}, which operates at 325~MHz and accelerates the beam to a kinetic energy of 2.5~MeV. Finally, a Medium Energy Beam Transport (MEBT) section matches the linear optics parameters of the accelerated beam to the injection section of IOTA. The parameters of the injected beam in IOTA are shown in Table~\ref{tab:protoninj}. The range of beam kinetic energies, $K_p$, depends on the accelerating fields inside the RFQ, while the bunch length, $\sigma_{z,\mathrm{inj}}$, and momentum spread, $\sigma_{\delta,\mathrm{inj}}$, are controlled by a separate 325~MHz de-bunching cavity \cite{GRomanov2011}. Based on previous studies \cite{Prebys2015}, we set the range of the momentum spread to $\sigma_{\delta,\mathrm{inj}} \in [1\times10^{-3}, 2\times10^{-3}]$. The transverse emittance of the proton beam is most sensitive to the configuration of the source and the LEBT, where the beam has low kinetic energy. Extrapolating previous measurements \cite{Tam2010} of a 50~keV proton beam using the same source and a similar LEBT configuration showed a normalized transverse emittance between $\approx 0.2 - 0.6$~\textmu m at a beam current of less than 8~mA, as illustrated in Fig.~\ref{fig:ipilebt}. Simulations suggest that the RFQ will contribute minimally to the emittance growth (see Table~6 in \cite{Ostroumov2006}). Therefore, we estimate a range of un-normalized rms emittance for the injected beam, considering the range of kinetic energies. Limiting the maximum normalized transverse emittance to 0.6~\textmu m in the LEBT constrains the current from the duoplasmatron source to $I_p \leq 8$~mA. Consequently, the maximum number of protons injected into IOTA is $I_p T_\mathrm{rev}/e = 9.13\times10^{10}$, where $T_\mathrm{rev}=1.83$~\textmu s is the nominal revolution time of a 2.5~MeV proton beam in IOTA. Both the emittance and the maximum current generated by the proton source depend on the diameter of the orifice expelling the beam from the plasma cup in the duoplasmatron, as well as various operational parameters. We will optimize the IPI configuration to adjust the beam parameters injected into IOTA for different experiments.\\

\section{Linear Optics in IOTA}

\begin{figure}
    \centering
    \includegraphics[width=0.75\columnwidth]{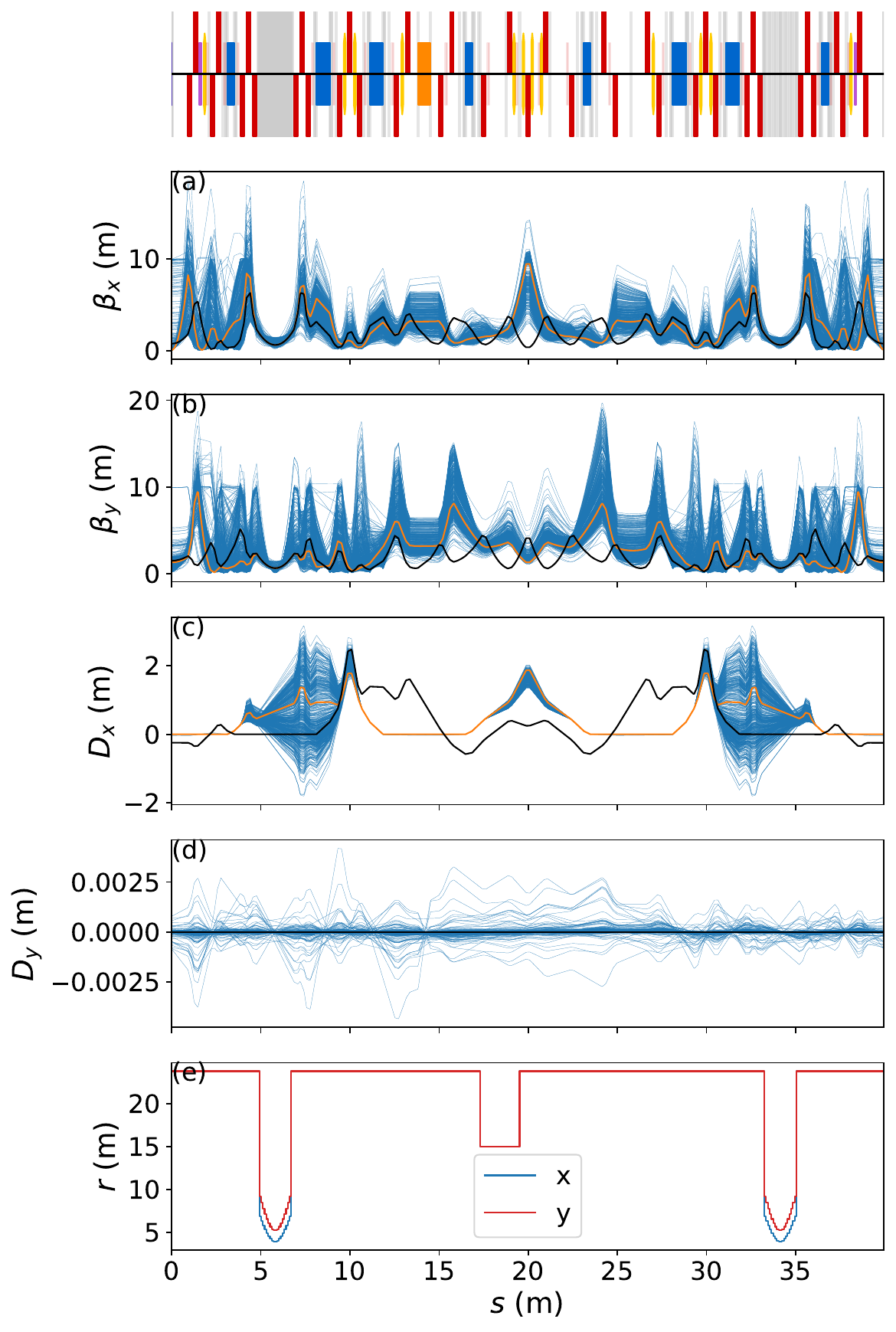}
    \caption{Twiss functions for two sets of lattice configurations. Panels (a) and (b) depict horizontal and vertical betatron amplitudes, while panels (c) and (d) display horizontal and vertical dispersions as functions of position, respectively. The black solid lines represent the \textit{DN} lattice, the orange lines denote the reference \textit{ECOOL} lattice, and the blue lines illustrate other members of the ensemble of \textit{ECOOL} lattices. Panel (e) shows the aperture radius as a function of position in the ring.}
    \label{fig:twiss}
\end{figure}

A wide range of linear optics configurations can be achieved in IOTA by adjusting the field strength of individual quadrupoles. A total of 79 free parameters allow us to control dispersion, one-turn phase advance, momentum compaction, and transverse coupling. Figure~\ref{fig:twiss} shows examples of betatron amplitudes ($\beta_{x,y}$) and dispersion ($D_{x,y}$) as functions of position ($s$) around the ring, computed using the MAD-X \cite{MADX} code, for two lattice configurations. The black curves represent the optics functions used for experiments with the Danilov-Nagaitsev (\textit{DN}) non-linear magnet \cite{Danilov2010}, the blue curves depict an ensemble of possible configurations for electron cooling (\textit{ECOOL}), and the orange curves indicate a reference lattice from this ensemble. \textaddition{The \textit{DN} lattice has already been implemented for 150~MeV electron operations in IOTA \cite{Romanov2014, Wieland2025} and benchmarked with linear optics calculations in MAD-X.} A separate document \cite{Banerjee2024-ecool} details the design of the electron cooling lattices \textaddition{for future use}. The transverse acceptances (\textcorrection{$\epsilon_{mx,y} \equiv \min { r_{x,y}^2/\beta_{x,y} }$}) of the lattice quantify the maximum phase-space area the beam can occupy and are dependent on the aperture radii ($r_{x,y}$). The non-linear magnets in sections BR and BL, as well as the planned Ionization Profile Monitor \cite{Piekarz2025} in section E, impose strict aperture limitations on the beam, significantly reducing the acceptance of the lattice. Table~\ref{tab:latticeparams} highlights this effect by comparing the acceptances with aperture restrictions to a configuration where the aperture is the nominal beam pipe radius of 25~mm throughout the ring. The transverse optics also influence the longitudinal dynamics of the ring.\\

We can configure IOTA to circulate either a coasting beam without longitudinal focusing or a bunched beam with the $h=4$ rf cavity \cite{Bruhaug2016} located in section DL, as shown in Fig.~\ref{fig:iota}. Since the 325~MHz rf system of the IPI is entirely independent of the 2.2~MHz cavity in IOTA, the injected beam rapidly debunches and fills the ring. The rf cavity can then adiabatically capture all the protons into buckets if the final bucket area exceeds the initial longitudinal phase-space area occupied by the coasting beam. This condition is satisfied if the equilibrium voltage exceeds a minimum value $V_\mathrm{min}$, given by \cite{Edwards1993}
\begin{equation}
    \label{eq:vmin}
    V_\mathrm{min}=\frac{\pi^3h|\eta|\beta^2\gamma m_p c^2 n^2 \sigma_{\delta,\mathrm{inj}}^2}{8e}\,,
\end{equation}
where $\eta$ is the phase slip factor, and $n$ is the number of standard deviations of the injected momentum spread to be captured. $\beta$ and $\gamma$ are the relativistic velocity and energy parameters, respectively. The maximum value of rf voltage is constrained by the available rf power driving the cavity, which is estimated to be 1~kV. In the bunched beam configuration, the steady-state rf voltage $V$ and the phase slip factor determine the synchrotron tune $Q_s$, \textcorrection{and the equilibrium momentum spread $\sigma_{\delta,\mathrm{eq}}$, which in turn determines the} bunch length $\sigma_s$ as follows, \cite{Wiedemann2015}
\begin{subequations}
    \begin{align}
        \label{eq:qs}
        Q_s &= \sqrt{\frac{h|\eta|}{2\pi\beta^2\gamma}\frac{eV}{m_pc^2}}\,, \\
        \beta_s &\equiv \frac{|\eta|C}{2\pi Q_s}\,, \\
        \sigma_s &= \beta_s \sigma_{\delta,\mathrm{eq}}\,,
    \end{align}
\end{subequations}
where $C$ is the circumference of the ring, and $\beta_s$ is defined as the longitudinal betatron amplitude. The range of cavity voltage\footnote{We set the minimum voltage so that particles with a momentum deviation of up to $2\sigma_{\delta,\mathrm{inj}}$ are captured.}, synchrotron tunes, and bunch lengths for the given range of injected beam energy, momentum spread, and phase slip factors corresponding to different lattices are shown in Table~\ref{tab:latticeparams}.

\begin{figure}
    \centering
    \includegraphics[width=0.45\columnwidth]{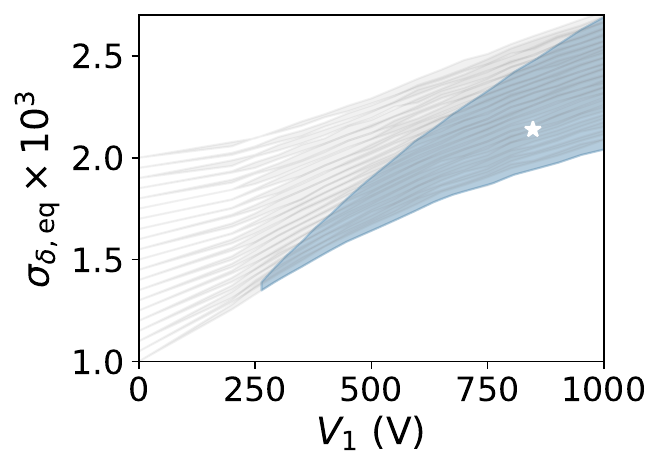}
    \caption{Equilibrium momentum spread ($\sigma_{\delta,\mathrm{eq}}$) as a function of final rf voltage ($V_1$) in the \textit{DN} lattice after completing the adiabatic capture process. Each gray band provides the range of $\sigma_{\delta,\mathrm{eq}}$ as a function of $V_1$ corresponding to the span of injected beam energies. The blue shaded region encompasses all configurations where the final RF voltage exceeds the minimum capture voltage. The star denotes the reference values of capture voltage and final momentum spread.}
    \label{fig:sigma_delta}
\end{figure}

We can obtain the equilibrium momentum spread $\sigma_{\delta,\mathrm{eq}}$ as a function of the initial momentum spread $\sigma_{\delta,\mathrm{inj}}$ of the injected beam using numerical simulations. We start with a coasting beam that fills the entire circumference of the IOTA ring and then adiabatically ramp the rf voltage $V(t)$ as a function of turns $t$ using \cite{Ng2002}
\begin{equation}
    V(t) = \frac{V_0}{\Big[1 - \big(1-\sqrt{\frac{V_0}{V_1}} \big)\frac{t}{t_1}\Big]^2}\,,
\end{equation}
where $V_1$ is the final rf voltage in equilibrium, and $t_1$ is the number of turns used to ramp the rf voltage. \textcorrection{The starting voltage $V_0$ must be non‑zero in order to keep the instantaneous synchrotron frequency finite and to satisfy the adiabaticity requirement (the relative change of the synchrotron‑bucket area is kept much smaller than the synchrotron frequency throughout the ramp). Its value is determined by the chosen adiabaticity factor $n_\mathrm{ad}$	and can be made arbitrarily small:}
\begin{equation}
    V_0 = \frac{V_1}{\Big[ 1+\frac{2\pi Q_s t_1}{n_\mathrm{ad}} \Big]^2}\,,
\end{equation}
where $Q_s$ is the equilibrium synchrotron tune corresponding to $V_1$, evaluated using Eq.~(\ref{eq:qs})\textcorrection{.} The results of numerical simulations conducted using the PyORBIT code \cite{Shishlo2015, PyORBIT-IOTA2024} are shown in Fig.~\ref{fig:sigma_delta}. In these simulations, we initialize a coasting beam modeled using 10\textsuperscript{5} macro-particles, with the nominal energy and transverse emittances specified in Table~\ref{tab:protoninj}, and track the beam for 2000~turns in the \textit{DN} lattice, \textaddition{accounting for single-particle dynamics only}. The adiabatic capture parameters are set to $t_1 = 1000$ and $n_\mathrm{ad} = 10$. We repeat these simulations for multiple values of injected momentum spread, beam energy, and final voltage, within the ranges given in Table~\ref{tab:protoninj}. Each gray band in Fig.~\ref{fig:sigma_delta} depicts $\sigma_{\delta,\mathrm{eq}}$ as a function of $V_1$ starting from a fixed value of $\sigma_{\delta,\mathrm{inj}}$. The lower and upper limits of the band corresponds to beam energies 2.5~MeV and 2.3~MeV respectively. The blue shaded region includes all configurations where \textcorrection{$V_1 \geq V_\mathrm{min}(2\sigma_{\delta,\mathrm{inj}})$, where the minimum voltage is calculated using Eq.~(\ref{eq:vmin}) in order to capture twice the injected rms momentum spread.} \textaddition{Since the initial momentum distribution is Gaussian, beam loss is less than 5~\% inside the blue shaded region.} Both the \textit{DN} lattice and the ensemble of \textit{ECOOL} lattices exhibit similar values of final momentum spread since the phase-slip factor for all lattices is $|\eta| \sim 1$. Consequently, the range of equilibrium momentum spread shown in Table~\ref{tab:latticeparams} is similar for both the \textit{DN} and the \textit{ECOOL} lattices.\\

The primary focus of research with protons in IOTA is to analyze the dynamics of beams with incoherent betatron tune shifts approaching 0.5. The beam distribution, intensity, and the geometry of the vacuum chamber determine the incoherent tune shifts. Assuming a beam with a Gaussian distribution in transverse phase-space, with rms size much smaller than the beam pipe, and a total intensity $N$, the incoherent tune shifts $\Delta\nu_{x,y,\mathrm{SC}}$ are, \cite{Litvinenko2015}
\begin{subequations}
    \begin{align}
        \label{eq:dnu}
        \Delta \nu_{x,y,\mathrm{SC}} &= \frac{\mathrm{d}N}{\mathrm{d}s}\frac{r_p}{2\pi\beta^2\gamma^3} \int_0^C \frac{\beta_{x,y}(s)}{\{\sigma_x(s)+\sigma_y(s)\}\sigma_{x,y}(s)}\,\mathrm{d}s\,,\\
        \label{eq:sigmaxy}
        \sigma_{x,y}(s) &= \sqrt{\beta_{x,y}(s)\epsilon_{x,y}+\sigma_\delta^2D_{x,y}^2(s)}\,,\\
        \label{eq:peaklinedensity}
        \frac{\mathrm{d}N}{\mathrm{d}s} &=
        \begin{cases}
            \frac{N}{C} & \text{for coasting}\\
            \frac{N}{\sqrt{2\pi}h\sigma_s} & \text{for bunched}\\
        \end{cases}
    \end{align}
\end{subequations}
where $\sigma_{x,y}(s)$, $\mathrm{d}N/\mathrm{d}s$ are the transverse beam sizes, peak line density of the beam, and $r_p$ is the classical radius of the proton. \textcorrection{The incoherent tune shift generally reduces with time as the emittance of the beam grows due to various heating mechanisms.} \textaddition{For instance, doubling the transverse emittance leads to the incoherent tune shift reducing by a factor of 2 as a consequence.} Here, we estimate the tune shifts only at the initial conditions of the beam. In the case of coasting beams, we use the values of injected transverse emittances and momentum spread listed in Table~\ref{tab:protoninj}, while in the case of bunched beams, we keep the same value of transverse emittance but use the values of relative momentum spread and bunch length after adiabatic capture. Table~\ref{tab:latticeparams} shows the range of beam intensity required to reach the intense space-charge regime for two lattices, taking into account the full range of energy, emittance, and momentum spread. The primary challenge to actually realizing such high tune shifts in the low-energy regime is the presence of various mechanisms that inflate the phase space and limit the storage time.\\

\begin{landscape}
\begin{table}
    \caption{Parameters for \textit{DN} and \textit{ECOOL} lattice configurations. Values in $()$ correspond to the nominal conditions, while the ranges are evaluated over space of beam parameters given in Table~\ref{tab:protoninj}. The time-scales associated with residual gas scattering correspond to a mean vacuum pressure of $10^{-9}$~Torr. The emittance growth rates (or reduction when rates are negative) due to intra-beam scattering and space-charge correspond to a beam intensity resulting in an initial vertical incoherent tune-shift of 0.5. $10^{10}$ protons in the ring correspond to a beam current of 0.88~mA at 2.5~MeV.}
    \label{tab:latticeparams}
    \centering
    \begin{tabular}{lccccl}
    \toprule
    \textbf{Parameter} & \multicolumn{2}{c}{\textbf{DN}} & \multicolumn{2}{c}{\textbf{ECOOL}} & \textbf{Unit} \\
    \midrule
    Betatron tune x ($Q_x$) & \multicolumn{2}{c}{5.30} & \multicolumn{2}{c}{$4.01 - 4.26$}\\
    Betatron tune y ($Q_y$) & \multicolumn{2}{c}{5.30} & \multicolumn{2}{c}{$3.01 - 3.51$}\\
    Acceptances with restrictions ($\epsilon_{mx},\epsilon_{my}$) & \multicolumn{2}{c}{$23,\, 41$}  & \multicolumn{2}{c}{$9.7 - 24\,(22),\, 9.1 - 42\,(40)$} & \textmu m \\
    Acceptances with nominal pipe ($\epsilon_{mx},\epsilon_{my}$) & \multicolumn{2}{c}{$90,\, 110$}  & \multicolumn{2}{c}{$31 - 87\,(60),\, 29 - 79\,(60)$} & \textmu m \\
    \midrule
    $\tau_\mathrm{RGS,single}$ with restrictions & \multicolumn{2}{c}{$93 - 4900\,(430)$} & \multicolumn{2}{c}{$16 - 4400\,(320)$} & s\\
    $\tau_\mathrm{RGS,single}$ with nominal pipe & \multicolumn{2}{c}{$320 - 17000\,(1500)$} & \multicolumn{2}{c}{$50 - 11000\,(670)$} & s\\
    $\tau_\mathrm{RGS,x}$ & \multicolumn{2}{c}{$4.1 - 560\,(27)$} & \multicolumn{2}{c}{$2 - 590\,(21)$} & s\\
    $\tau_\mathrm{RGS,y}$ & \multicolumn{2}{c}{$4.4 - 610\,(20)$} & \multicolumn{2}{c}{$1.6 - 510\,(15)$} & s\\
    \midrule
    & Coasting & Bunched & Coasting & Bunched \\
    \midrule
    Harmonic number ($h$) & --- & 4 & --- & 4\\
    Cavity voltage ($V$) & --- & $0.26 - 1\,(0.85)$ & --- & $0.24 - 1\,(0.84)$ & kV\\
    Synchrotron tune ($Q_s$) & --- & $0.006 - 0.011\,(0.01)$ & --- & $0.005 - 0.012\,(0.01)$ \\
    RMS momentum spread ($\sigma_{\delta,\mathrm{eq}}$) & $1 - 2\,(1.3)$ & $1.4 - 2.7\,(2.1)$ & $1 - 2\,(1.3)$ & $1.3 - 2.6\,(2.1)$ & $10^{-3}$\\
    RMS bunch length ($\sigma_s$) & --- & $0.70 - 2.8\,(1.3)$ & --- & $0.64 - 2.7\,(1.3)$ & m \\
    Intensity ($|\delta\nu_{y,\mathrm{SC}}|=0.5$) & $6.2 - 20\,(7.4)$ & $1.1 - 7.1\,(2.4)$ & $5.2 - 21\,(7.3)$ & $0.92 - 7.0\,(2.4)$ & $10^{10}$\\
    \midrule
    $\tau^{-1}_\mathrm{IBS,x}$ & $-0.02 - 0.38\,(0.14)$ & $0.00 - 1.08\,(0.50)$ & $-0.05 - 0.91\,(0.21)$ & $-0.09 - 2.50\,(0.68)$ & s\textsuperscript{-1}\\
    $\tau^{-1}_\mathrm{IBS,y}$ & $-0.08 - 0.22\,(0.13)$ & $-0.15 - 0.68\,(0.58)$ & $-0.08 - 0.62\,(0.24)$ & $-0.14 - 1.72\,(0.86)$ & s\textsuperscript{-1}\\
    $\tau^{-1}_\mathrm{IBS,z}$ & $-0.02 - 0.85\,(0.19)$ & $-0.12 - 0.93\,(-0.05)$ & $-0.05 - 0.68\,(0.11)$ & $-0.14 - 0.84\,(-0.07)$ & s\textsuperscript{-1}\\
    \midrule
    $\tau_\mathrm{SC,x}$ & $(89)$ & $(7.3)$ & $(22)$ & $(8.9)$ & ms\\
    $\tau_\mathrm{SC,y}$ & $(41)$ & $(6.0)$ & $(16)$ & $(6.7)$ & ms \\
    $\tau_\mathrm{SC,z}$ & --- & --- & --- & --- & \\
    \bottomrule
    \end{tabular}
\end{table}
\end{landscape}

\section{Emittance growth and beam lifetime due to residual gas scattering}

\begin{figure}
    \centering
    \includegraphics[width=0.5\linewidth]{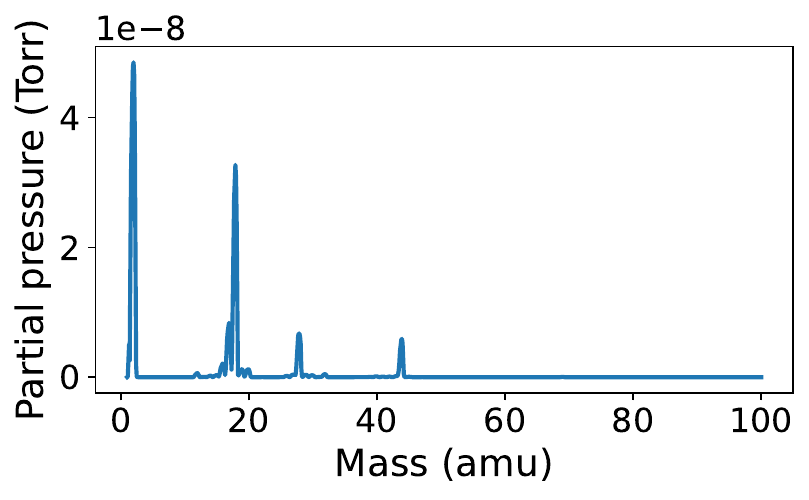}
    \caption{Partial pressure of residual gas as a function of molecular weight as measured by a Residual Gas Analyzer mounted in IOTA.}
    \label{fig:rgascan}
\end{figure}

\begin{table}
    \caption{Mole fraction $\chi \equiv P/P_\mathrm{tot}$ of different molecules in the IOTA vacuum chamber for three different scenarios.}
    \label{tab:vacuum}
    \centering
    \begin{tabular}{llccc}
        \toprule
        \textbf{Gas} & \textbf{Mass (amu)} & \textbf{Light} & \textbf{Reference} & \textbf{Heavy} \\
        \midrule
        H\textsubscript{2} & 2 & 1.0 & 0.507 & 0.0 \\
        NH\textsubscript{3} & 17 & 0.0 & 0.086 & 0.0 \\
        H\textsubscript{2}O & 18 & 0.0 & 0.296 & 0.0 \\
        CO & 28 & 0.0 & 0.062 & 0.0 \\
        CO\textsubscript{2} & 44 & 0.0 & 0.049 & 1.0 \\
        \midrule
        Total & & 1.0 & 1.0 & 1.0 \\
        \bottomrule
    \end{tabular} 
\end{table}

Coulomb scattering of protons by the nuclei of residual gas atoms inside the vacuum system strongly contributes to the emittance growth of the beam and is the primary mechanism of beam loss at low intensities. The frequency and distribution of impact parameters of the scattering events depends on the composition of the residual gas inside the vacuum chamber. Figure~\ref{fig:rgascan} shows the partial pressures ($P_j$) of different molecules in the IOTA vacuum chamber, measured using a Residual Gas Analyzer (RGA). We use the position of the peaks in the mass spectrum to identify 5~major constituents, and tabulate their mole fractions in Table~\ref{tab:vacuum} under the designation \textit{Reference}. We use two other compositions containing only hydrogen, and only carbon dioxide, designated \textit{Light} and \textit{Heavy} respectively. Assuming that these compositions are uniform along the ring, we can estimate nominal values and ranges for rates of emittance growth and beam loss.\\

\begin{figure}
    \centering
    \includegraphics[width=0.9\columnwidth]{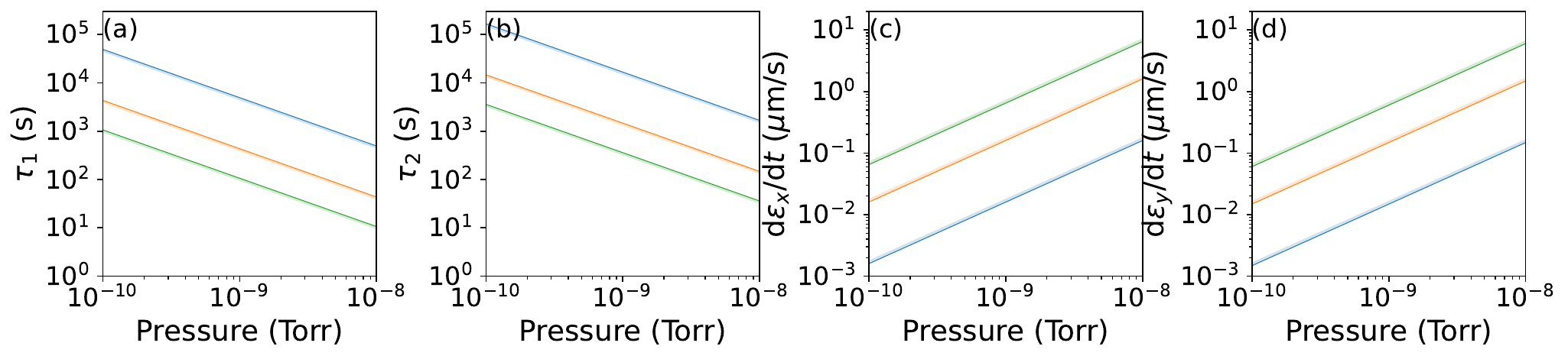}
    \includegraphics[width=0.9\columnwidth]{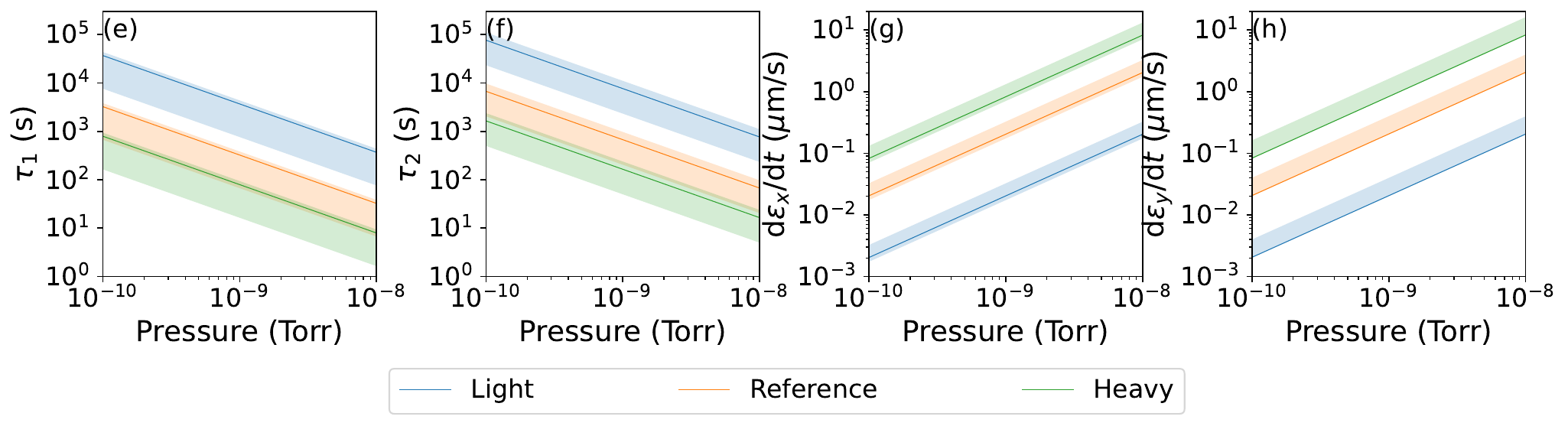}
    \caption{Range of beam loss and emittance growth timescales for different vacuum scenarios and lattices. Panels (a) and (b) illustrate the beam lifetime in the \textit{DN} lattice with and without aperture restrictions, respectively. Panels (c) and (d) depict the horizontal and vertical emittance growth rates for the \textit{DN} lattice, respectively. Panels (e) and (f) present the beam lifetime for the \textit{ECOOL} lattices with and without aperture restrictions, respectively. Lastly, panels (g) and (h) show the horizontal and vertical emittance growth rates for the \textit{ECOOL} lattices, respectively. The shaded regions represent the range of values evaluated over the space of beam parameters and lattice configurations\textcorrection{, while the solid lines denote the nominal values.}}
    \label{fig:rgs}
\end{figure}

Multiple \textcorrection{inelastic scattering events with atomic electrons} increase the angular distribution of the beam, leading to steady growth of transverse emittance ($\mathrm{d}\epsilon_{x,y}/\mathrm{d}t$), given by (Combining Eq.~(10) and (11) from Sec. 3.3.2 in \cite{Handbook2023}),
\begin{equation}
    \frac{\mathrm{d}\epsilon_{x,y}}{\mathrm{d}t} = \frac{2\pi cr_p^2}{\beta^3\gamma^2} \langle \beta_{x,y} \rangle \sum_i n_i Z_i (Z_i+1) L_{c,i}\,,
\end{equation}
where $\langle \beta_{x,y} \rangle$ represents the transverse betatron amplitudes averaged over the ring. The emittance growth rate depends on the number density ($n_i$) of each nucleus with charge ($Z_i$) and also on the Coulomb logarithm $L_{c,i} \approx \ln 183/Z_i^{1/3}$, which encodes the range of impact parameters of the scattering events. \footnote{We use the partial pressures of different gas molecules tabulated in Table~\ref{tab:vacuum} to evaluate the number density of specific atoms in the vacuum system. For instance, the total number density of oxygen atoms is related to the mole fractions of all gas molecules containing oxygen atoms, i.e., $n_\text{O} = P_\text{tot}(\chi_{\text{H}_2\text{O}}+\chi_\text{CO}+2\chi_{\text{C}\text{O}_2})/(kT)$, where $P_{tot} = \sum_j P_j$ is the total vacuum pressure, $T$ is the room temperature, and $k$ is the Boltzmann constant.} Apart from the gas composition, the growth rate also depends on beam energy. Figure~\ref{fig:rgs} shows the emittance growth rates over a range of gas pressure, compositions, and beam energy, indicating a spread of almost two orders of magnitude at any given gas pressure, regardless of the lattice.\\

Single scattering events that trigger betatron oscillations with amplitudes exceeding the acceptance of the ring lead to beam loss. We can evaluate the beam lifetime ($\tau_\mathrm{RGS, single}$) as the inverse of the total scattering rate,
\begin{equation}
    \label{eq:rgslifetime}
    \tau^{-1}_\text{RGS,single} = \frac{2\pi cr_p^2}{\beta^3\gamma^2} \bigg(\frac{\langle \beta_x \rangle}{\epsilon_{mx}} +\frac{\langle \beta_y \rangle}{\epsilon_{my}} \bigg) \frac{P_\text{eff}}{kT}\,,
\end{equation}
where we define an effective pressure $P_\text{eff} \equiv kT \sum_i n_i Z_i (Z_i + 1)$ measured in atomic hydrogen equivalent. Panels (a) and (b) in Fig.~\ref{fig:rgs} show the range of beam lifetimes for the \textit{DN} lattice as a function of total pressure with and without the aperture restrictions, respectively. Panels (e) and (f) show the same for the \textit{ECOOL} family of lattices. Table~\ref{tab:latticeparams} shows the lifetimes corresponding to a mean pressure of $10^{-9}$~Torr in the ring, illustrating that removing the aperture restrictions can improve the lifetime by a factor of two or more depending on the lattice. While the loss rate depends on the gas pressure, composition, linear optics setup, and beam energy, it is independent of beam emittance and intensity, as long as it is small compared to the aperture.\\

We can improve beam lifetime and emittance growth rates by baking the IOTA vacuum chamber \cite{Thompson2025} to reduce moisture content. Starting from the \textit{reference} gas composition, we can estimate the relative change in effective pressure as follows:
\begin{equation}
    \label{eq:deltapeff}
    \frac{\Delta P_\text{eff}}{P_\text{eff}} = \frac{ \{ 1(1+1)2+8(8+1)1\} \Delta \chi_{\text{H}_2\text{O}}}{
        \splitfrac{\{1(1+1)2\}\chi_{\text{H}_2}+\{7(7+1)1+1(1+1)3\}\chi_{\text{N}\text{H}_3}+\{ 1(1+1)2+8(8+1)1\} \chi_{\text{H}_2\text{O}}}
        {+\{6(6+1)1+8(8+1)1\}\chi_{\text{CO}}+\{6(6+1)1+8(8+1)1\}\chi_{\text{CO}_2}}
    } \approx 1.651 \Delta \chi_{\text{H}_2\text{O}} \,,
\end{equation}
where $\Delta \chi_{\text{H}_2\text{O}}$ is the change in the mole fraction of water after baking. Combining this result with Eq.~(\ref{eq:rgslifetime}), a 100-fold reduction in the partial pressure of water is estimated to boost the beam lifetime by a factor of 2.\\

The planned installation of the Ionization Profile Monitor (IPM) in section E also introduces considerations regarding residual gas content within the IOTA vacuum system. The IPM relies on the controlled injection of noble gases, such as Krypton or Xenon, into a localized section to enhance ion production and, consequently, the collected signal. The design incorporates multiple ion pumps and apertures to provide isolation and minimize the population of additional gas atoms within the remainder of the IOTA vacuum system. Introducing noble gases with high atomic numbers, coupled with the added apertures, may lead to increased emittance growth rates and a reduction in beam lifetime. A more detailed discussion of these issues can be found in separate documentation. \cite{Piekarz2025}\\

\section{Emittance growth due to intra-beam scattering}
Elastic scattering between protons inside the beam leads to momentum transfer between different degrees of freedom and an overall decrease in the phase-space density. \textaddition{To estimate the magnitude of this effect, we employ a simple model for the growth rates of the transverse emittance and the relative momentum spread caused by intra‑beam scattering (IBS).} \textcorrection{Within the smooth‑lattice approximation\textemdash assuming zero dispersion and a round beam\textemdash the growth rates can be expressed as}, \cite{Nagaitsev2005}
\begin{subequations}
    \label{eq:ibs}
    \begin{align}
        \frac{\mathrm{d} \epsilon_\perp}{\mathrm{d} t} &= A_\perp \frac{\mathrm{d}N}{\mathrm{d}s} \frac{r_p^2 c L_C}{\beta^3 \gamma^5 \epsilon_\perp \theta_\perp} F(z)\,, \\
        \frac{\mathrm{d} \sigma_\delta^2}{\mathrm{d} t} &= A_\parallel \frac{\mathrm{d}N}{\mathrm{d}s} \frac{r_p^2 c L_C}{\beta^3 \gamma^3 \sigma_\perp^2 \theta_\perp} F(z)\,, \\
        F(z) &\approx (1-z^{1/4}) \frac{\ln (z+1)}{z}\,, \\
        z &= \frac{\sigma_\delta^2}{\gamma^2\theta_\perp^2}\,,
    \end{align}
\end{subequations}
where $\epsilon_\perp$, $\sigma_\perp \equiv \sqrt{\epsilon_\perp \beta_\perp}$, $\theta_\perp \equiv \sqrt{\epsilon_\perp/\beta_\perp}$, and $\sigma_\delta$ are the rms transverse emittance, beam size, beam divergence, and relative longitudinal momentum spread, respectively. The dimensionless parameter $z$ determines the direction of thermal energy transfer between the transverse and longitudinal degrees of freedom. Specifically, $z<1$ implies that there is less thermal energy in the longitudinal degree of freedom with respect to the transverse, and consequently, $\mathrm{d}\epsilon_\perp/\mathrm{d}t<0$, while the flow reverses for $z>1$. The values of peak line density of the protons are given in Eq.~(\ref{eq:peaklinedensity}) while the pre-factors $A_\perp$ and $A_\parallel$ are given by,
\begin{subequations}
    \begin{align}
        A_\perp &=
        \begin{cases}
            -\frac{\sqrt{\pi}}{8} & \text{for coasting}\\
            -\frac{\sqrt{\pi}}{8\sqrt{2}} & \text{for bunched}\\
        \end{cases}\,,\\
        A_\parallel &=
        \begin{cases}
            \frac{\sqrt{\pi}}{2} & \text{for coasting}\\
            \frac{\sqrt{\pi}}{4\sqrt{2}} & \text{for bunched}\\
        \end{cases}\,.
    \end{align}
\end{subequations}
For a 2.5~MeV proton beam, with $\epsilon_\perp = 4$~\textmu m, $\sigma_\delta = 10^{-3}$, $\sigma_s = 1$~m, circulating inside IOTA with $\beta_\perp=2$~m, we find that $z\approx0.5$, which implies that relative momentum spread increases while the transverse emittance shrinks. Assuming a typical value of the Coulomb logarithm $L_C \sim 20$ \cite{Bjorken1982, Reiser2008}, yields longitudinal growth time-scales of 3.1 and 7.7~seconds for coasting and bunched beams, respectively. However, these simple estimates of IBS rates ignore the transverse anisotropy of phase-space and do not account for realistic linear optics functions of the lattice.\\

\begin{figure}
    \centering
    \includegraphics[width=0.9\columnwidth]{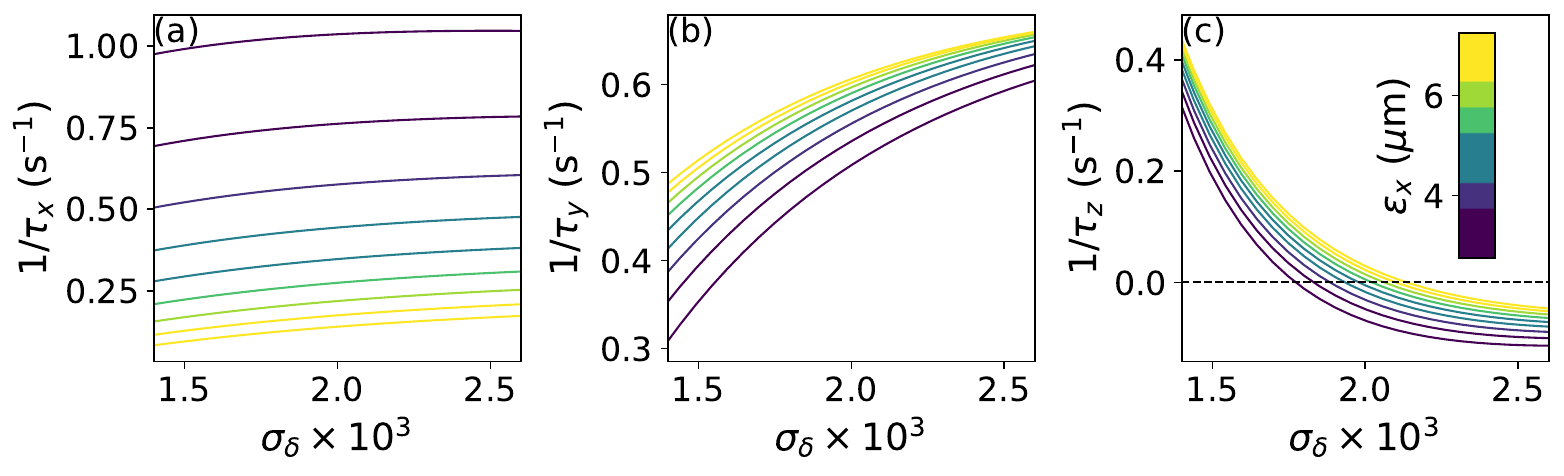}
    \caption{Growth rates of (a) horizontal emittance, (b) vertical emittance, and (c) momentum spread, driven by intra-beam scattering as functions of momentum spread ($\sigma_\delta$) and horizontal rms emittance ($\epsilon_x$) in the \textit{DN} lattice with bunched beam. The beam energy and vertical emittance are fixed to the nominal values listed in Table~\ref{tab:protoninj}.}
    \label{fig:ibstest}
\end{figure}

We use the IBS module \cite{Antoniou2012} in the MAD-X code for a comprehensive estimation of emittance growth rates for the full range of beam parameters. We plot the dependence of IBS growth rates $1/\tau_\mathrm{IBS,x,y} \equiv (1/
\epsilon_{x,y}) \mathrm{d}\epsilon_{x,y}/\mathrm{d}t$ and $1/\tau_\mathrm{IBS,z} \equiv (1/\sigma_\delta^2) \mathrm{d}\sigma_\delta^2/\mathrm{d}t$ in bunched beam as functions of the horizontal emittance and the relative momentum spread in Fig.~\ref{fig:ibstest}, while keeping the beam energy and the vertical emittance at their nominal values. We observe that the longitudinal growth rate changes sign in panel (c) at a threshold value of $\sigma_\delta$, which is consistent with the dependence suggested in the simple model depicted in Eq.~(\ref{eq:ibs}); however, a similar change of sign is not observed in the transverse degrees of freedom. In general, the higher the thermal energy in a particular degree of freedom, the lower its growth rate.\\

\begin{figure}
    \centering
    \includegraphics[width=0.8\columnwidth]{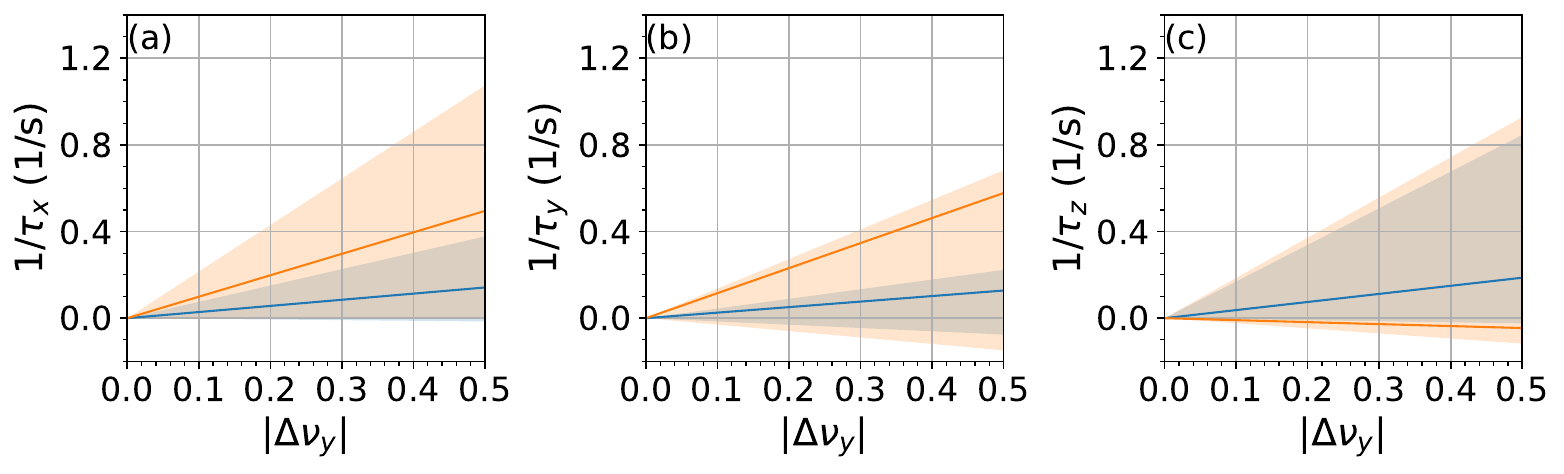}
    \includegraphics[width=0.8\columnwidth]{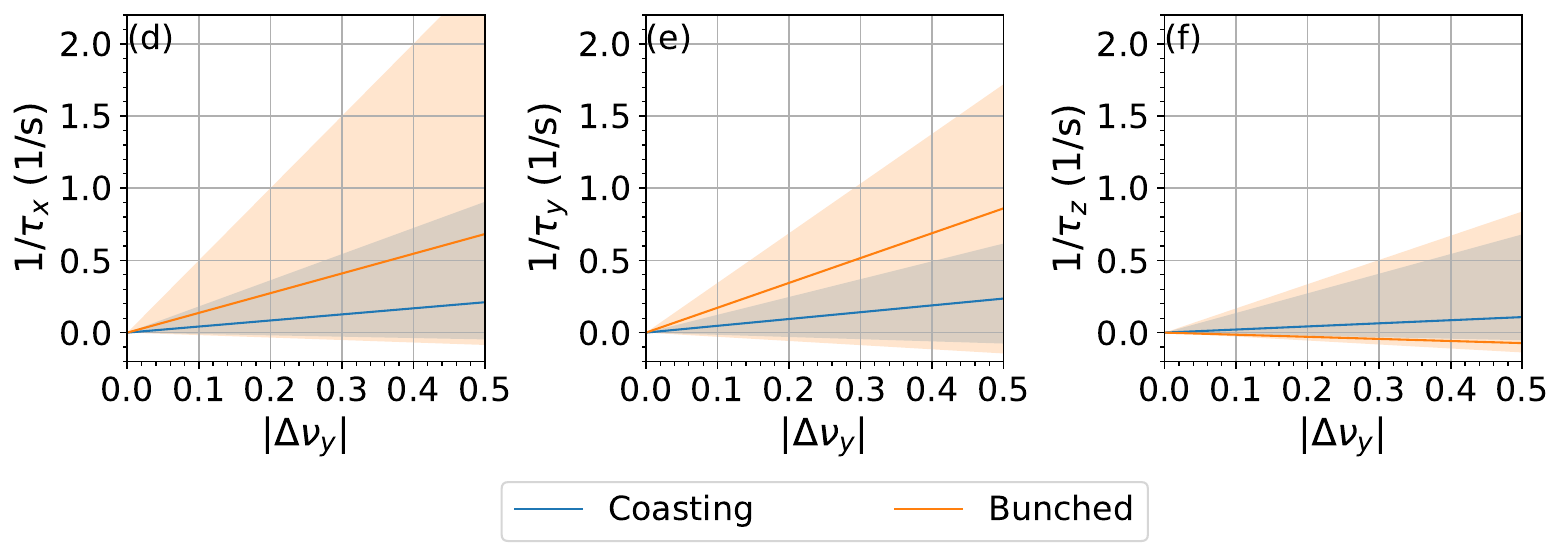}
    \caption{Range of emittance growth rates as functions of vertical space-charge tune shift for \textit{DN} (panels (a), (b), (c)) and \textit{ECOOL} (panels (d), (e), (f)) lattices. The shaded regions represent the range, while the solid lines denote the nominal values.}
    \label{fig:ibs}
\end{figure}

The emittance growth rates due to IBS vary monotonously over the entire parameter space of beam energy, bunch length, transverse emittances and momentum spread. Hence, to estimate the range of growth rates, we calculate IBS rates at the boundaries of the 5-dimensional parameter space and aggregate the results. Figure~\ref{fig:ibs} shows such an estimation for both coasting and bunched beams as functions of the vertical space-charge tune shift. Depending on the exact configuration, IBS can either very quickly expand or even lead to a slow reduction in thermal energy at the expense of a different degree of freedom. The solid lines in Fig.~\ref{fig:ibs} represent nominal beam parameters with the reference lattices and can be used as a guide while commissioning. We list the range of emittance growth rates and the nominal values corresponding to $|\delta\nu_{y,\mathrm{SC}}|=0.5$ in Table~\ref{tab:latticeparams}.\\

\section{Space-Charge Dynamics in bare IOTA}
Apart from short-range interactions due to elastic scattering, collective effects driven by space-charge and wakefields contribute to beam dynamics at high intensities. In the case of IOTA proton operations, wakefields generated by the interaction of the beam with the vacuum chamber are negligible, while space-charge effects dominate. The transverse rms beam size ($\sigma_{x,y}$) relative to the Debye length ($\lambda_{D,p}$) of the beam core dictates whether coherent effects or \textcorrection{incoherent single-particle effects} govern the evolution of phase-space. We define the Debye length in the center-of-mass frame, at the core of a Gaussian beam as follows,
\begin{subequations}
    \begin{align}
        \label{eq:np}
        n_p &= \frac{\mathrm{d}N}{\mathrm{d}s} \frac{1}{2\pi\gamma\sigma_x\sigma_y}\,,\\
        \label{eq:vth}
        v_\mathrm{th} &= \gamma\beta c\sqrt{\frac{\epsilon_x}{\beta_x}+\frac{\epsilon_y}{\beta_y}+\frac{\sigma_\delta^2}{\gamma^2}}\,,\\
        \label{eq:debye}
        \lambda_{D,p} &= \frac{v_\mathrm{th}}{\omega_p}\,,
    \end{align}
\end{subequations}
where $n_p$, $v_\mathrm{th}$ and $\omega_p=\sqrt{n_p e^2/(m_p \epsilon_0)}$ are the number density, rms thermal velocity, and plasma frequency of protons at the beam core, respectively. In addition, we can define a characteristic length scale of separation between protons in the beam frame as $d_p \equiv (4\pi n_p/3)^{-1/3}$ and a length scale associated with the period of plasma oscillation $l_p \equiv 2\pi c \beta\gamma/\omega_p$ in the lab frame\textaddition{, both at the core of the beam}. The range of the characteristic length scales, for both coasting and bunched beams at the incoherent tune shift $|\delta\nu_{y,\mathrm{SC}}|=0.5$, in the \textit{DN} and \textit{ECOOL} lattices over the full range of beam parameters, including energy, emittances, momentum spread and bunch length are shown in Table~\ref{tab:scparams}. Based on these ranges, we can conclude that the Debye length is comparable or larger than the transverse size of the bunch. Thus, single-particle resonances such as those from magnetic field multipoles or from space-charge modulations (structure resonances) drive emittance growth and beam loss. \cite{Hofmann2017} We use the Particle-in-Cell (PIC) simulation code PyORBIT to model the evolution of phase-space in the IOTA lattice.\\

\begin{table}
    \caption{Characteristic scales for space-charge dynamics in IOTA for an intense beam with tune shift $|\delta\nu_{y,\mathrm{SC}}|=0.5$, along with parameters of the 2.5~D space-charge model used in PyORBIT.}
    \label{tab:scparams}
    \centering
    \begin{tabular}{lccl}
        \toprule
        \textbf{Parameter} & \textbf{Nominal} & \textbf{Range} & \textbf{Unit} \\
        \midrule
        RMS transverse beam sizes ($\sigma_x,\sigma_y$) & --- & $0.38 - 13$ & mm \\
        Debye length ($\lambda_{D,p}$) & --- & $3.8 - 21$ & mm \\
        Characteristic separation ($d_p$) & --- & $6.5 - 46$ & \textmu m \\
        Plasma oscillation period ($l_p$) & --- & $3.5 - 64$ & m \\
        \midrule
        Transverse grid size ($\Delta x_\mathrm{PIC} = \Delta y_\mathrm{PIC}$) & $50/128 \approx 0.39$ & --- & mm \\
        Longitudinal grid resolution & 128 & --- \\ 
        Distance between kicks ($\Delta s_\mathrm{PIC}$) & --- & $\leq 0.20$ & m \\
        Number of macro-particles ($N_\mathrm{macro}$) & $10^5$ & --- \\
        \bottomrule
    \end{tabular}
\end{table}

We choose the 2.5~D PIC space charge model in PyORBIT, which applies transverse kicks to all macro-particles at various positions in the lattice, assuming that the number density profile of individual transverse slices of the bunch remain constant over its longitudinal extent. \textcorrection{Table~\ref{tab:scparams} shows that the transverse grid spacing is set to the pipe diameter divided by the chosen grid resolution. This choice satisfies two constraints: the cell size is larger than the typical inter‑proton spacing, yet smaller than both the Debye length and the overall beam size.} \textaddition{With a total of $10^5$ macro-particles spread over a transverse grid of $128^2$, cells near the beam core have at least 6 macro-particles each.} Conversely, the longitudinal grid has a fixed resolution while the extent is dynamically adapted to the positions of the macro-particles. The maximum distance between space-charge kicks ($\Delta s_\mathrm{PIC}$) is much smaller than the characteristic length of a plasma oscillation. Further, we choose the number of macro-particles to exceed the total number of cells in the 2~D grid, and thus obtain rms values of phase-space variables with relative uncertainties at the few-percent level. To be consistent with the calculations in the previous sections, we use \textaddition{linear} elements of bare IOTA (only dipoles and quadrupoles) with no errors, no multipole fields\textcorrection{, an uniform aperture diameter of 50~mm everywhere, and open boundary conditions for the space-charge field.} We initialize the beam with a Gaussian distribution in transverse phase-space and either a uniform or Gaussian distribution in longitudinal space depending on whether the beam is coasting or bunched, respectively.\\

\begin{figure}
    \centering
    \includegraphics[width=\linewidth]{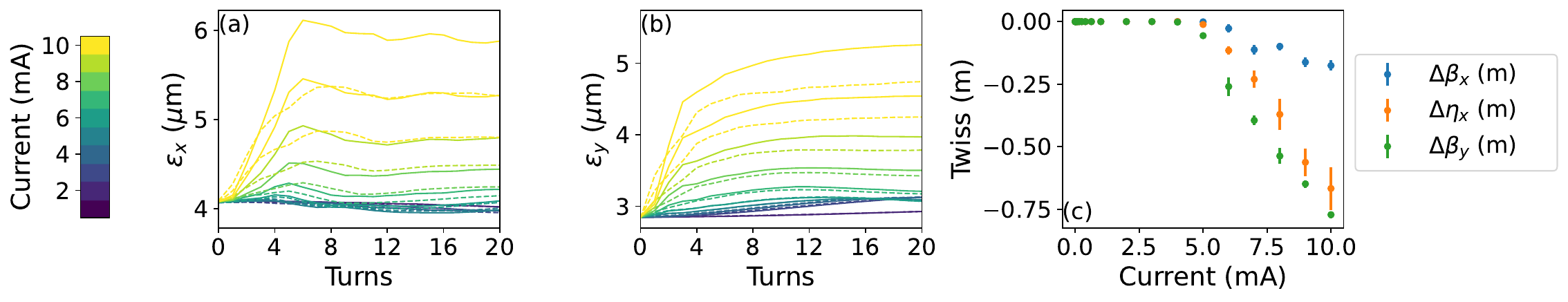}
    \caption{Emittance growth of coasting beam driven by betatron mismatch due to space-charge in the \textit{DN} lattice. Panels (a) and (b) show the horizontal and vertical emittance growth, respectively, in IOTA over 20~turns after injection. The solid lines represent emittance growth when starting with Twiss parameters of the bare lattice \textaddition{(including linear elements without space-charge de-focusing)}, while the dashed line shows the same when starting with optimized Twiss parameters. Panel (c) shows the difference of optimum Twiss parameters from the bare values as a function of beam current.}
    \label{fig:scmatchcoasting}
\end{figure}

As the injected beam relaxes into an equilibrium distribution over the first few turns, the mechanism of initial emittance growth is the mismatch between the initial beam distribution and the effective focusing created by the lattice and the beam’s self-field. To minimize this mismatch-driven emittance growth, we optimized the Twiss parameters of the injected distribution. Specifically, we defined a loss function based on the transverse emittance growth after 20 simulated turns and used the Broyden–Fletcher–Goldfarb–Shanno algorithm to search over $\beta_x$, $\beta_y$, and $\eta_x$ (with all other Twiss functions set to zero since the lattice is mirror-symmetric about the center). Figure~\ref{fig:scmatchcoasting} shows the results of this optimization for coasting-beam injection into the \textit{DN} lattice as a function of beam current. As seen in panels (a) and (b), the optimization yields up to a 10~\% reduction in final transverse emittance at 10~mA, with smaller gains at lower currents. The optimal Twiss values in panel (c) may serve as a guide when commissioning IOTA. \\

\begin{figure}
    \centering
    \includegraphics[width=\linewidth]{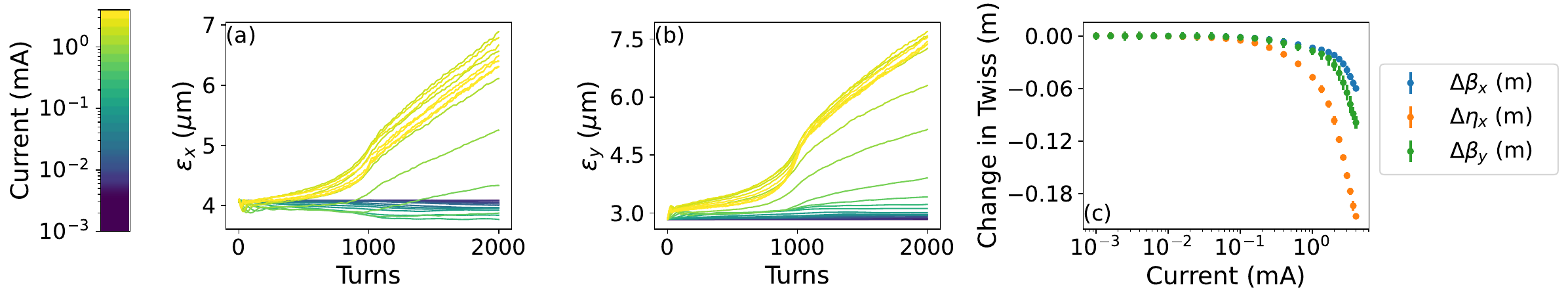}
    \caption{Emittance growth due to space-charge during the adiabatic capture process with the \textit{DN} lattice for a range of injected beam currents. Panels (a) and (b) show the horizontal and vertical emittance growth, respectively while panel (c) shows the change in Twiss parameters measured at 2000~turns after injection, compared to initial values, on beam current.}
    \label{fig:scmatchbunched}
\end{figure}

A second phase of emittance growth takes place during the adiabatic capture process which bunches the beam thereby increasing the charge-density at the beam-core creating a mismatch. The resulting emittance growth and equilibrium Twiss functions as seen at the injection location of the lattice are shown in Fig.~\ref{fig:scmatchbunched}. While the entire bunching process takes place over the first 1000~turns, the period between 700 to 1000~turns displays pronounced emittance growth, followed by steady, but slower growth after. The dependence of equilibrium Twiss functions on beam current, as measured at the end of 2000~turns, is similar to that of the optimum Twiss functions in Fig.~\ref{fig:scmatchcoasting}(c). The steady emittance growth, seen after adiabatic bunching is complete is caused by individual particles crossing structure resonances.\\

\begin{figure}
    \centering
    \includegraphics[width=\linewidth]{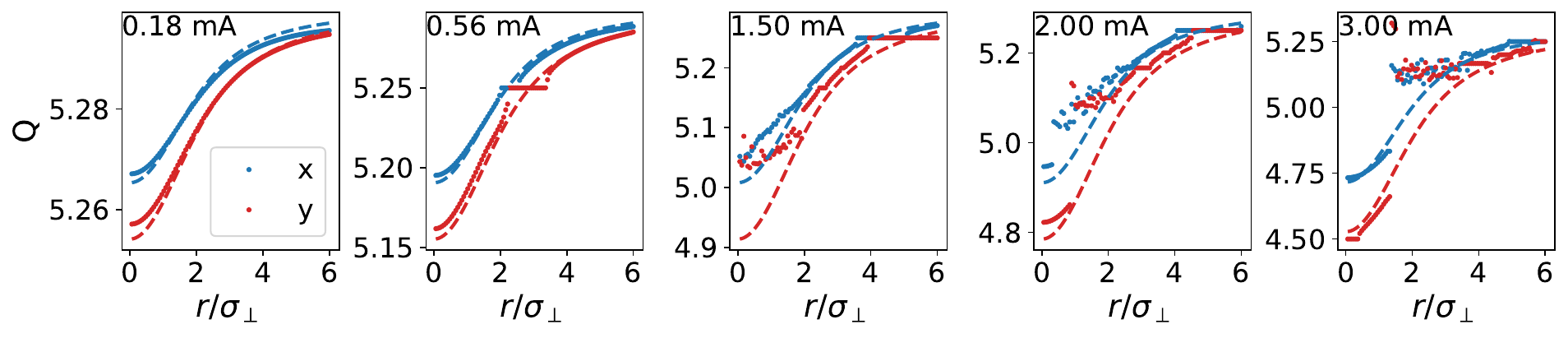}
    \caption{Transverse betatron tunes $Q_{x,y}$ of test particles as functions of initial amplitude $r/\sigma_\perp$ in the \textit{DN} lattice for a set of 5~selected bunched beam currents, where $\sigma_\perp$ is the rms transverse beam size in either the horizontal or vertical direction. Dots represent tunes measured from simulation while the dashed lines are results from Eq.~(\ref{eq:dnuxax}). Flat areas and discontinuities indicate resonance trapping and crossing respectively.}
    \label{fig:scresonances}
\end{figure}

Particles experiencing periodic kicks from magnetic multipoles or from space-charge forces undergo resonance excitation which is a principal contributor to steady emittance growth and halo formation in storage rings. In the absence of space-charge forces, the tune footprint is dominated by the chromatic tune spread and amplitude-dependent detuning generated by non-linear magnets such as octupoles. The intrinsic tune spread at zero current is widened by a contribution from space-charge forces, which reduces the effective focusing seen by particles near the core of the beam. We can use an analytical model \cite{Sen2023} to estimate the tune shift of particles inside a Gaussian bunch as a function of normalized amplitude as follows,
\begin{equation}
    \label{eq:dnuxax}
        \Delta\nu_x (a_x) \approx \frac{\Delta \nu_{x,\mathrm{SC}}}{\big\langle \frac{2}{\sigma_y/\sigma_x + 1} \big\rangle_s} \int^1_0 \exp\bigg[ -\frac{a_x^2 u}{4} \bigg] \Bigg[ I_0 \bigg( \frac{a_x^2 u}{4} \bigg) - I_1 \bigg( \frac{a_x^2 u}{4} \bigg) \Bigg] \bigg\langle \bigg[ \frac{1}{(\sigma_y^2/\sigma_x^2-1)u + 1} \bigg]^{1/2} \bigg\rangle_s \mathrm{d}u\,,
\end{equation}
where $a_x \equiv x/\sigma_x$ is the normalized amplitude of a test particle started with a horizontal position $x$, with all other phase-space coordinates set to 0. $\langle \rangle_s$ represents the average value along the ring. $I_0$ and $I_1$ are the \nth{0} and \nth{1} order modified Bessel functions of the first kind, respectively. We obtain the vertical tune shift by exchanging $x \longleftrightarrow y$ in the above expression. This model agrees well with PyORBIT simulations using a frozen space-charge model \cite{Bartosik2019}, at a relatively low beam current of 0.18~mA, as seen in the leftmost panel of Fig.~\ref{fig:scresonances}. As we increase the beam current, we observe particles being trapped at resonances or crossing them entirely appearing as discontinuities in the plots. The beam crosses more resonances as we increase beam current. Specifically for the  \textit{DN} lattice with bunched beam, the \nth{4}, \nth{6}, integer and half-integer resonances are crossed sequentially with increasing beam current. For bunched beam configurations, synchrotron motion drive periodic resonance crossing \cite{Franchetti2003}, while in coasting configurations, an interplay of space-charge structural instabilities and resonances \cite{Hofmann2015} result in steady emittance growth and beam loss.\\

\begin{figure}
    \centering
    \includegraphics[width=0.8\linewidth]{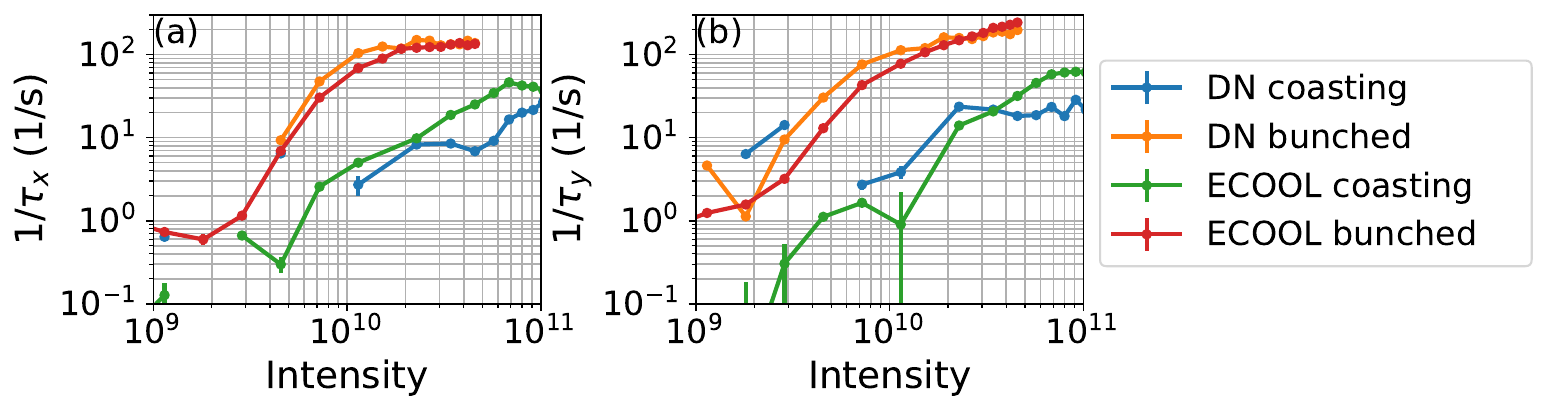}
    \caption{Transverse emittance growth rates due to space-charge in the \text{DN} and the reference \textit{ECOOL} lattices, as functions of beam current with other parameters set to their reference values. Panels (a) and (b) show the horizontal and vertical growth rates respectively.}
    \label{fig:scgrowth}
\end{figure}

We estimated emittance growth rates for both the \textit{DN} and the reference \textit{ECOOL} lattices for nominal beam parameters, in both bunched and coasting beam configurations using self-consistent PIC simulations in PyORBIT. Panels (a) and (b) of Fig.~\ref{fig:scgrowth} show the growth rates of rms horizontal and vertical emittance, respectively measured at 2000~turns after injection. The emittance growth time-scales are of the order of 10~ms at space-charge tune shifts approaching 0.5, compared to time-scales closer to 1~s purely due to IBS. Table~\ref{tab:latticeparams} shows approximate time scales associated with space-charge related emittance growth. It is important to note that forces evolve self-consistently with changes in emittance and beam intensity, and gradually become smaller as the beam size increases.\\

\begin{figure}
    \centering
    \includegraphics[width=0.8\linewidth]{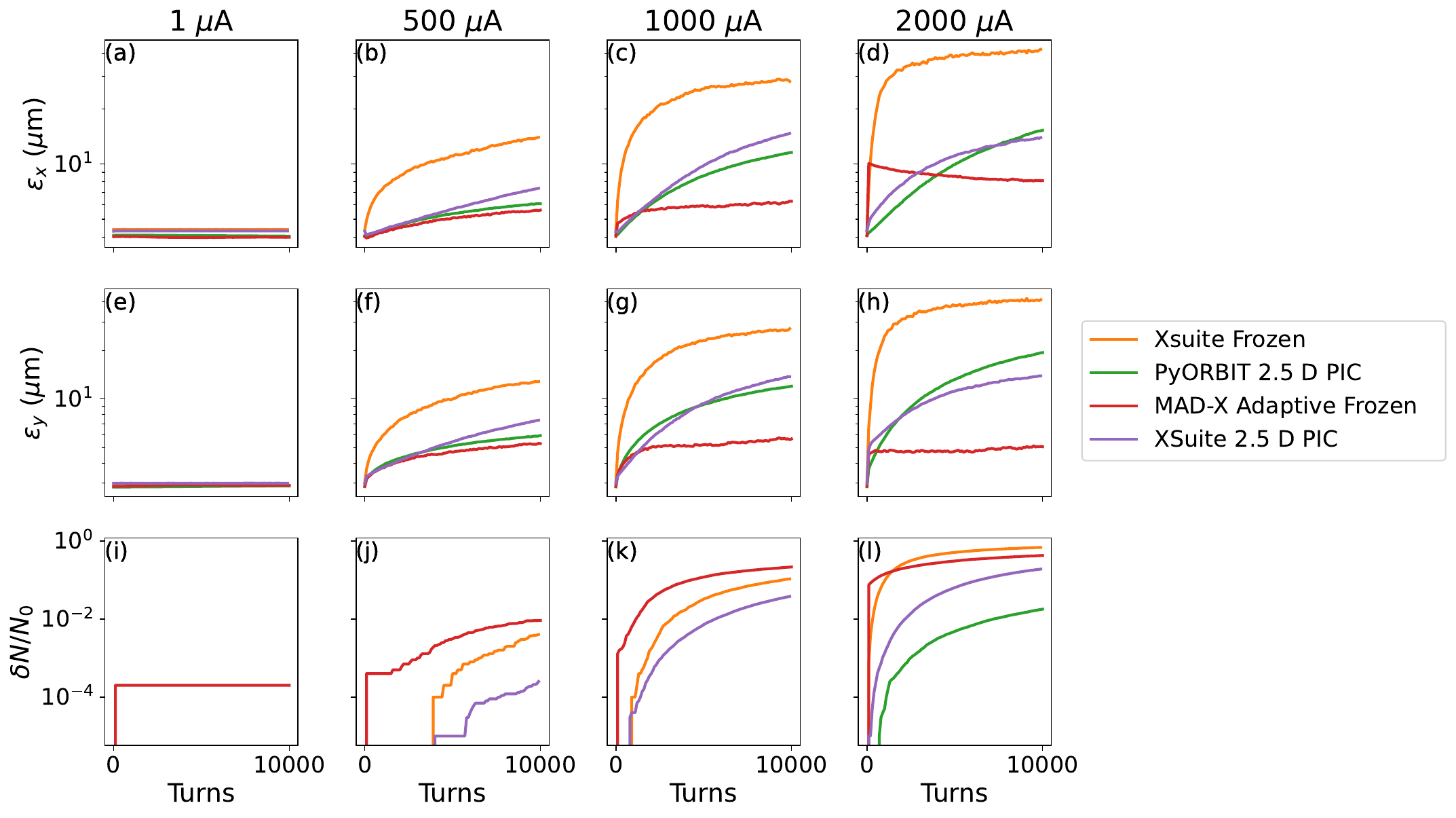}
    \caption{Comparison of transverse rms emittance and beam loss in the \textit{DN} lattice, as a function of turn number for different simulation models. Panels (a), (b), (c), and (d) show the evolution of horizontal emittance, while panels (e), (f), (g), and (h) illustrate vertical emittance. Panels (i), (j), (k), and (l) depict fractional beam loss. The four columns represent beam currents of 1~\textmu A, 0.5~mA, 1~mA, and 2~mA, respectively.}
    \label{fig:scevolution}
\end{figure}

\begin{figure}
    \centering
    \includegraphics[width=0.8\linewidth]{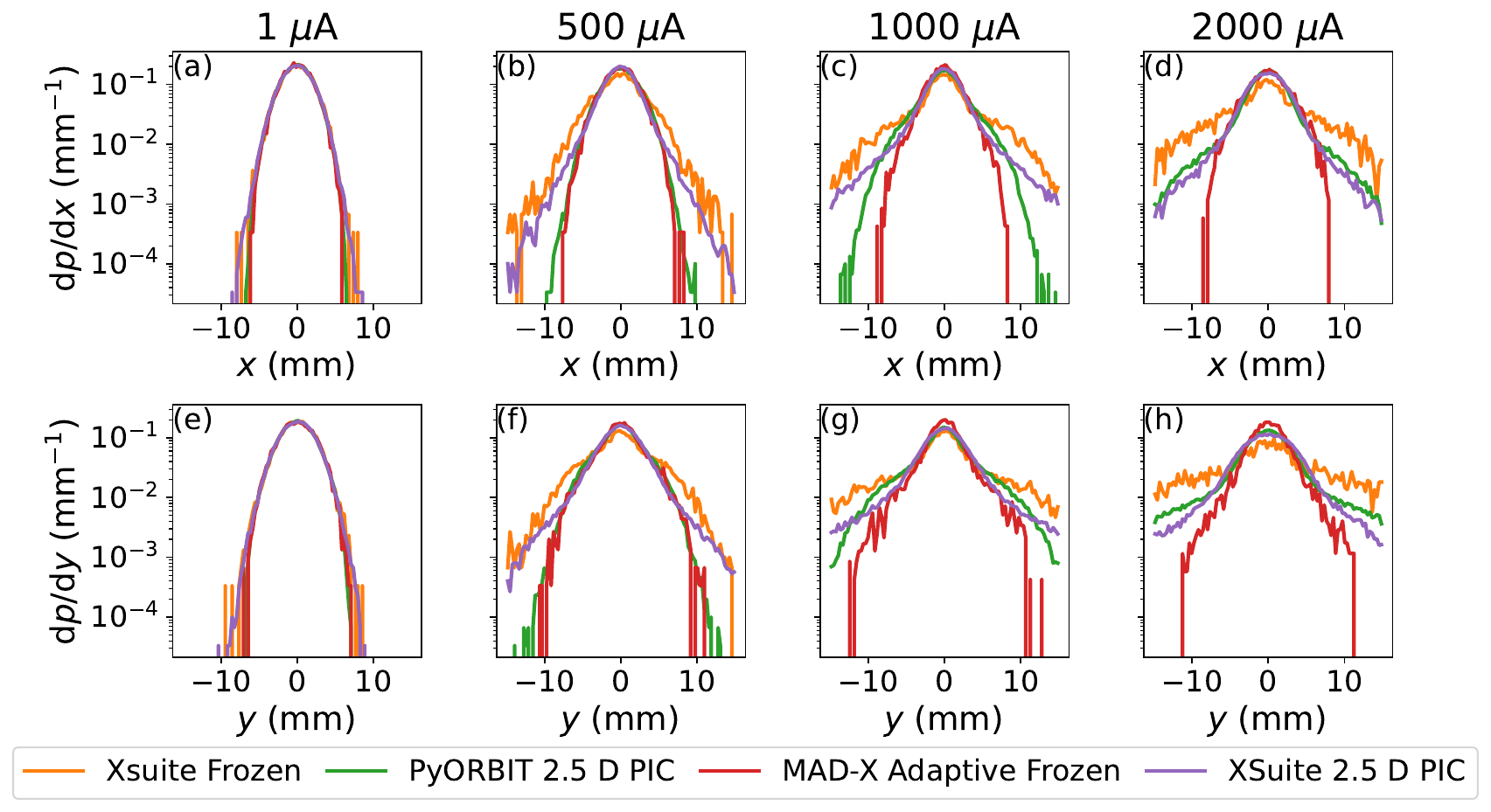}
    \caption{Comparison of transverse density profiles for a bunched beam in IOTA after completing $10^4$~turns, corresponding to the simulation results in Fig.~\ref{fig:scevolution}. Panels (a), (b), (c), and (d) display horizontal distributions, while panels (e), (f), (g), and (h) illustrate vertical distributions.}
    \label{fig:sc1dhistograms}
\end{figure}

We illustrate the long-term evolution of bunched beams at four intensities \cite{Banerjee2024} in the \textit{DN} lattice, using four simulation models: a 2.5~D transverse PIC model in PyORBIT, a similar 2.5~D PIC model in XSuite \cite{Iadarola2023}, a frozen model in XSuite, and an adaptive-frozen model \cite{Schmidt2016} in MAD-X.\footnote{\textcorrection{Both frozen models represent the three‑dimensional charge distribution as a transverse 2‑D Gaussian multiplied by a longitudinal 1‑D Gaussian. The transverse electric field of the 2‑D Gaussian is obtained from the Bassetti‑Erskine formula \cite{Bassetti1980} and then weighted by the local longitudinal charge density to calculate the transverse kicks. Both frozen models separate the 3D spatial distribution into a transverse 2D Gaussian weighted with a longitudinal 1D Gaussian.}} Figure~\ref{fig:scevolution} compares rms emittance and beam loss, while Fig.~\ref{fig:sc1dhistograms} compares the transverse profiles after $10^4$~turns. At a current of 1~\textmu A, the transverse incoherent tune shift due to space charge is $\sim 2\times10^{-4}$, leading to linear dynamics where the phase-space distribution remains Gaussian and emittance growth is minimal, consistent across all \textcorrection{four models}. At moderate beam currents  corresponding to an incoherent tune shift of less than 0.3, significant increases in transverse rms emittance are observed in all models, albeit with some quantitative differences.\\

While the emittance evolution in the self-consistent PIC models in XSuite and PyORBIT are similar, the transverse beam distributions and the beam loss are different. Specifically, the results from the XSuite PIC model indicates the formation of more intense halo as seen in Fig.~\ref{fig:sc1dhistograms} compared to the PyORBIT model, and hence more beam loss. \textaddition{This discrepancy can be traced to two implementation differences: (i) PyORBIT is configured to use a fixed transverse grid, whereas the XSuite PIC model adapts the grid to the instantaneous rms beam size; and (ii) the number and longitudinal positions of the space‑charge kicks differ between the codes (both sets were individually converged, but the discretization is not identical). The frozen calculations also diverge. In the XSuite frozen model the transverse emittance is assumed constant, so the transverse beam size remains fixed with time even as the actual emittance grows; consequently the space‑charge force does not weaken as the beam blows up, leading to a faster emittance rise. The MAD‑X adaptive‑frozen model updates the beam size (and thus the space‑charge fields) turn‑by‑turn, which reduces the field strength as the beam expands and moderates the emittance growth. Moreover, MAD‑X applies both transverse and longitudinal apertures at the exit of every element, while XSuite is configured to impose a single transverse aperture at the ring entrance.} Once the incoherent tune shift exceeds 0.3, particles at the center of the beam cross the integer resonance, resulting in rapid beam blow-up, with all four simulations diverging in emittance growth and beam loss.\\

\textaddition{The above discussion provides a starting point for understanding these differences, and a detailed benchmarking study of the various models using the IOTA 2.5~MeV proton‑beam dynamics scenario will be presented in future work (see Ref. \cite{Schmidt2016} for previous benchmarking efforts involving MAD‑X and PyORBIT).} These simulations \textaddition{also} neglect realistic magnetic field errors and multipole fields which can excite further resonances and lead to faster emittance growth and beam loss.\\

\section{Discussion and Conclusions}
\begin{figure}
    \centering
    \includegraphics[width=0.8\linewidth]{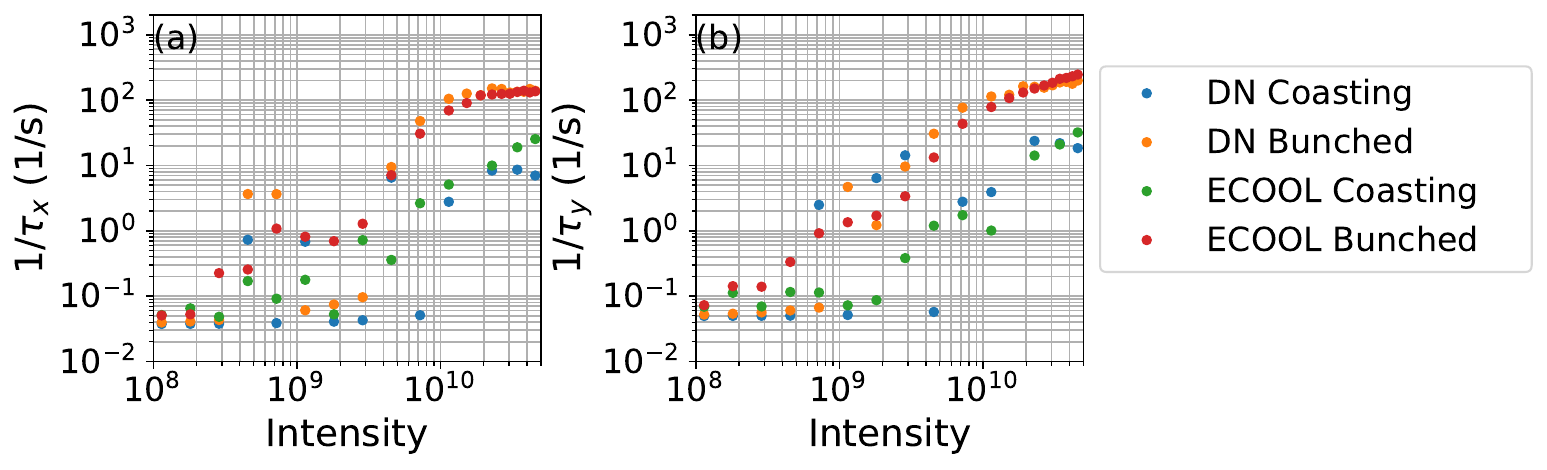}
    \caption{Net growth rates of transverse emittance due to RGS, IBS and space-charge in the \text{DN} and the reference \textit{ECOOL} lattices, as functions of beam current with other parameters set to their reference values. Panels (a) and (b) show the horizontal and vertical growth rates respectively.}
    \label{fig:netgrowth}
\end{figure}

The circulation of a 2.5~MeV proton beam in IOTA provides a unique platform to study the interplay of space-charge with non-linear dynamics and collective effects in a highly-flexible magnetic field configuration. However, the very low energy of the proton beam enhances intra-beam scattering, residual gas scattering, and space-charge-driven emittance growth and beam loss. We estimate the strength of all these effects in the absence of magnetic field errors and summarize the growth and loss rates in Table~\ref{tab:latticeparams}. Based on a recent measurement of the vacuum composition of the IOTA, single scattering events with residual nuclei limits the beam lifetime to 7~minutes or less at the nominal beam parameters due to the tight aperture constraints imposed by the Danilov-Nagaitsev non-linear magnet. Beyond this intensity-independent beam loss mechanism, emittance growth also leads to gradual beam loss as particles with large amplitudes scrape the vacuum chamber. Figure~\ref{fig:netgrowth} shows the net transverse emittance growth rates as functions of intensity for both the \textit{DN} and \textit{ECOOL} lattice configurations with coasting and bunched beam. At low intensities, $\lesssim 10^8$, the intensity-independent mechanism of multiple scattering with residual gas dominates, at intermediate intensities between $10^8$ to $10^9$, intra-beam scattering dominates while above $10^9$, space-charge starts dominating. Consequently, at high-intensities ($|\Delta \nu_{y,\mathrm{SC}}| \gtrsim 0.1$), storage times will be limited to $10^5$~turns or less which will constrain the design of experiments.\\

While this report provides order-of-magnitude estimates of emittance growth and beam loss, driven by various mechanisms, more detailed calculations will be required to reproduce actual measurements with decent precision. At intermediate tune-shifts ($|\Delta \nu_{y,\mathrm{SC}}| \in [0.01, 0.1]$), a precise estimate of intra-beam scattering will require accounting for the formation of non-Gaussian tails, \cite{Papadopoulou2020} which increase the rms size of the beam without changing the core density. Moreover, configurations with transverse coupling lead to a re-partitioning of emittance growth rates which requires a dedicated model \cite{Lebedev2018}. At high-intensities, space-charge dynamics is greatly affected by the initial beam distribution since it directly changes the incoherent tune distribution of the particles. Apart from the non-linearities arising from the beam distribution itself, magnetic field errors, and intentional non-linearities in the lattice also affect the dynamics. Hence realistic models of space-charge must include field maps of all magnets included with an ensemble of typical errors encountered during operation. Better estimates of beam loss with tracking simulations need to include accurate aperture specifications throughout the lattice. Such detailed calculations should be done based on the needs for individual experiments planned in IOTA.\\

We can develop measures to mitigate the driving mechanisms of emittance growth and beam loss for various experiments. Baking the IOTA vacuum chamber above $100\,^\circ$C can reduce the residual gas content seen in Fig.~\ref{fig:rgascan}, especially the partial pressure of water, leading to a better beam lifetime and reduction in emittance growth rates at low intensities. At intermediate intensities, incorporating an electron cooler \cite{Banerjee2024-ecool} as a part of the electron lens program can compensate for IBS and lead to better control of the phase-space of the beam. At high-intensities where space-charge plays an important role, optimization of the bare tune \textaddition{(linear optics without space-charge)}, inclusion of periodicity and compensation of Resonance Driving Terms \cite{Rabusov2022, Gonzalez-Ortiz2024} can reduce emittance growth. For instance, compensating for the \nth{4} order resonance in the \textit{DN} lattice by turning on octupoles can greatly reduce resonance-driven diffusion at tune shifts $|\Delta \nu_{y,\mathrm{SC}}| \lesssim 0.3$ while keeping the linear optics unchanged. Beside conventional measures, the IOTA research program will also explore NIO to improve stability at resonances, and space-charge compensation using an electron lens \cite{Shiltsev2017, Stancari2021, Nagaitsev2021} to reduce the effective tune footprint of the beam with broad applicability in high-intensity synchrotrons and storage rings.\\

We explored the available scenarios for experiments with protons in IOTA. Proton dynamics dictates the ranges of achievable lifetimes, loss rates and emittance growth rates. A wide range of experiments is planned, such as nonlinear integrable optics with space charge, control of instabilities with feedback systems, and the interplay between lattice nonlinearities, tunable impedances and space charge. The studies presented in this paper provide the foundations and tools for the design and interpretation of experiments in the upcoming IOTA experimental campaigns. \\

\section*{Acknowledgements}
We would like to thank Rob Ainsworth, Sergey Antipov, Alexey Burov, Brandon Cathey, Valery Lebedev, Chad Mitchell, Sergei Nagaitsev, Frank Schmidt, Tanaji Sen, Ben Simons, Eric Stern, Alexander Valishev, and John Wieland for insightful discussions, for help with modeling and simulations, and for contributing to the definition of the IOTA proton physics program. This work was produced by FermiForward Discovery Group, LLC under Contract No.~89243024CSC000002 with the U.S.~Department of Energy, Office of Science, Office of High Energy Physics. Publisher acknowledges the U.S. Government license to provide public access under the DOE Public Access Plan DOE Public Access Plan.

\end{document}